\documentclass[12pt,preprint]{aastex}

\begin{document}

\title{Seyfert 2 Galaxies with Spectropolarimetric Observations}

\author{Qiusheng Gu and Jiehao Huang}

\affil{Department of Astronomy, Nanjing University, Nanjing
210093, P. R. China}

\email{qsgu,jhh@nju.edu.cn}

\begin{abstract}
We present a compilation of radio, infrared, optical and hard
X-ray (2-10kev) data for a sample of 90 Seyfert 2 galaxies(Sy2s)
with spectropolarimetric observations (41 Sy2s with detection of
polarized broad lines  (PBL) and 49 without PBL). Compared to Sy2s
without PBL, Sy2s with PBL tend to be earlier-type spirals, and
show warmer mid-infrared color and significant excess of emissions
(including the hard X-ray(2-10kev), [O {\sc iii}]$\lambda$5007,
infrared (25 $\mu$m) and radio). Our analyses indicate that the
majority of Sy2s without PBL are those sources having less
powerful AGN activities, most likely caused by low accretion rate.
It implies that the detectability of the polarized broad emission
lines in Sy2s may depend on their central AGN activities in most
cases. Based on the available data, we find no compelling evidence
for the presence of two types of Sy2s, one type of them has been
proposed to be intrinsically different from Sy2s claimed in
Unification Model.
\end{abstract}


\section{Introduction}
In the scheme of the standard unification model, Seyfert 1 and 2
galaxies (Sy1s and Sy2s hereafter) are intrinsically the same
objects and the absence of broad emission lines in Sy2s is
ascribed to the obscuration by a pc-scale dusty torus oriented
along the line of sight (see the reviews by  Antonucci 1993). The
observational evidence for this model includes the detection of
polarized broad emission lines in some Seyfert 2 galaxies
(Antonucci \& Miller 1985; Tran 1995 and 2001; Young et al. 1996;
Heisler et al. 1997; and Moran et al. 2000), the detection of
broad lines in the infrared spectra of some Sy2s (Rix et al. 1990;
Ruiz et al. 1994; and Veilleux et al. 1997) and the detection of a
prominent photoelectric cutoff in the X-ray spectra of Sy2s
indicating the presence of large columns of gas along the line of
sight (Koyama et al. 1989; Awaki et al. 1991; Maiolino et al.
1998; Risaliti et al. 1999). However, recent investigations
suggest that this strictest version of unification model needs
modifications. Among them, we may find, for example, the
outflowing wind model (Elvis 2000), and the existence of two
intrinsically different populations of Sy2s $-$ the hidden Sy1 and
the "real" Sy2 with a weak or absent Sy1 nucleus $-$ ( Tran 2001)
based on a spectropolarimetric survey of the CfA  and 12$\mu$m
samples of Seyfert 2 galaxies. Nevertheless, Antonucci (2001)
strongly argued that the evidence claimed by Tran is quite
uncertain.

With the improvement of the techniques and instruments for the
spectropolarimetry, now people have observed a large sample of
Sy2s with less bias (e.g.  Heisler et al. 1997; Moran et al.
2000). For the present,  polarized broad lines (PBL) have been
detected in several dozens of  Sy2s, while not detected in other
several dozens. Although such surveys are probably biased
inherently since pre-selection was done according to the
broad-band polarization, they still provide a largest sample for
more meaningful analysis than before so that we may, or may not,
find some compelling evidence for the proposed modifications,
especially the presence of two types of Sy2s. That is what we
would present in this paper.

In this work, we provide the multiwavelength data for a sample of
90 Seyfert 2 galaxies collected from recent literatures in \S2,
and compare the properties of host galaxies, infrared, radio and
hard X-ray continua and [O {\sc iii}] emission in \S3. The
implications and discussions on the results are given in \S4, and
conclusions in \S5 .

\section{The Sample}
\label{sect:sample} We collect all Seyfert 2 galaxies with
available spectropolarimetric data from  recent literature (from
1985 to 2002), where 41 Sy2s with PBL and 49 without the detection
of PBL are found.  Just as  most,  if not all , Seyfert samples,
the current spectropolarimetry sample of Sy2s  (SPSS  hereafter)
is heterogeneous and cannot be considered complete, though it
covers a wide range of AGN luminosities  (about 4 orders of
magnitudes).

The infrared and radio data for these SPSS sources are listed in
Table 1, which contains galaxy name (column 1); the IRAS color
f$_{\rm 60}$/f$_{\rm 25}$ (column 2); the 25 $\mu$m luminosity
(column 3); the far-infrared (40 $-$ 120 $\mu$m) luminosity
(column 4), determined by (Lonsdale et al. 1985)
\[
L_{\rm FIR}=3.75\ 10^5 \ D^2\ (2.58 f_{\rm 60} + f_{\rm 100})
\]
where D is the distance in Mpc (the Hubble constant is taken as 75
km s$^{-1}$ Mpc$^{-1}$), f$_{\rm 60}$ and f$_{\rm 100}$ are the
flux densities at 60 $\mu$m and 100 $\mu$m in Jy; the radio powers
at 1.49 GHz (column 5) from the NRAO/VLA Sky Survey (NVSS) (Condon
et al. 1998); the mean revised morphological types (column 6) from
the Third Reference Catalogue of Bright Galaxies (RC3, de
Vaucouleurs et al. 1991); and  the  information for PBL (columns 7
and 8). Following Lumsden et al. (2001),  we adopt the infrared
fluxes from BGS (Soifer et al. 1989; Sanders et al. 1995) for Sy2s
with f$_{60} >$ 5Jy; from Strauss et al. (1990) for those with
f$_{60} <$ 5Jy, and IRAS Faint Source Catalog (FSC, Moshir et al.
1992) for all other ones.

In Table 2, we present the optical and hard X-ray information as
the following: column 1, the galaxy name; column 2, the blue
luminosity computed with the formula (Pogge \& Eskridge 1993)
\[ {\rm log} \ L_{\rm B} = 12.208 - 0.4B_{\rm T}^{0} + {\rm
log}(1+z) + 2 {\rm log} D \] where z is the redshift of the
galaxy; column 3, the luminosity of H$_{\beta}$ emission; column
4, the luminosity of extinction-corrected [O {\sc
iii}]$\lambda$5007 emission given as $\rm L_{\rm [O III]} = 4 \pi
D^2 F_{\rm [O III]}^{\rm cor}$, where F$_{\rm [O III]}^{\rm cor}$
is the extinction-corrected flux of [O {\sc iii}]$\lambda$5007
emission derived from the relation (Bassani et al. 1999) \[ F_{\rm
[O III]}^{\rm cor} = F_{\rm [O III]}^{\rm obs} \rm
(\frac{(H_\alpha/H_\beta)_{obs}}{(H_\alpha/H_\beta)_{0}})^{2.94}
\]
we assume an intrinsic Balmer decrement $\rm
(H_\alpha/H_\beta)_{0} = 3.0$; column 5, the references for
H$_\beta$ and [O {\sc iii}]$\lambda$5007; column 6, gaseous
absorbing column density (N$_{\rm H}$); column 7, the
absorption-corrected hard X-ray (2-10kev) luminosity for
Compton-thin Seyfert 2 galaxies; column 8, the corresponding
references; and column 9, Eddington luminosity  acquired  by $
L_{\rm Edd} = 1.51 \ 10^{38} M_{\rm BH}/M_{\odot} \ erg s^{-1} $.
The central black hole masses are estimated  by use of  the
M$_{\rm BH} - \sigma$ relation by Merritt \& Ferrarese (2000),
which is \[ M_{\rm BH} = 1.3 \times 10 ^{8} (\frac{\sigma}{200 \
km \ s^{-1}})^{4.72} \ M_{\odot}
\]
Nuclear velocity dispersions ($\sigma$) are taken from Nelson \&
Whittle (1995) and McElroy (1995) if  available.

\section{The results}

\subsection{Properties of host galaxies}

Following the approach taken by Storchi-Bergmann et al (2001) to
evaluate the morphological type distribution for their Seyfert
sample, we display the histogram distributions of morphological
types for our  Sy2s {\it with} and {\it without} PBL  (SPSS1 and
SPSS0 for simplicity hereafter) in Fig 1a and 1b, respectively.
The statistical median of their morphological type distributions
are Sa and Sab for SPSS1 and SPSS0, showing with stars in Fig1a
and 1b, respectively. The SPSS1 (Sy2s {\it with} PBL) obviously
tend to be earlier-type spirals.

In Fig.2, we present the cumulative distributions of blue
luminosities (in unit of L$_\odot$) for SPSS1 and SPSS0, denoted
with filled and open circles, respectively.  The two-sample
Kolmogorov-Smirnov (KS) test, {\it kolmov} task  in
IRAF\footnote{IRAF is distributed by the National Optical
Astronomy Observatories, which are operated by the Association of
Universities for Research in Astronomy, Inc., under cooperative
agreement with the National Science Foundation.}, shows that there
exists no significant difference between the two subsamples. The
same conclusion could be reached with the Kaplan-Meier estimator,
{\it kmestimate} task in IRAF, yielding L$_{\rm B}$ of 10.475
$\pm$ 0.066 and 10.349 $\pm$ 0.085 for SPSS1 and SPSS0,
respectively. The basic statistical results presented in this
paper are summarized in Table 3.

We also compare the column densities (N$_{\rm H}$) for SPSS1 and
SPSS0. Due to the existence of censored data in column densities
(20 sources with lower limits and 1 source with upper limit, NGC
7590), the survival analysis methods (ASURV Rev 1.2
\footnote{http://www.astro.psu.edu/statcodes/asurv}, Isobe \&
Feigelson, 1990) has been used for the statistical analysis.  A
possibility of  79\% for the test is given by the Gehan's
Generalized Wilcoxon Test - Hypergeometric Variance (GGW test, one
kind of {\it asurv} tests), suggesting little difference in $\log
N_{\rm H}$ distributions between SPSS1 and SPSS0. Following
Alexander (2001) and Tran (2001), we may also estimate the mean
$\log  N_{\rm H}$ of 23.63 $\pm$ 0.20 cm$^{-2}$ and 23.56 $\pm$
0.35 cm$^{-2}$  (23.67 $\pm$ 0.35 cm$^{-2}$ if NGC 7590 is
excluded, since {\it asurv} could not deal with the case which
contains both upper and lower limits) for the two subsamples,
respectively,  which leads to the same conclusion on the column
density (N$_{\rm H}$). Indeed, Alexander (2001), Gu et al. (2001)
and Tran (2001) have derived the similar results for samples with
smaller size.

\subsection{Indicators of AGN activities} During the last two
decades, there have been several attempts to determine which
emissions are truly isotropic for Seyfert galaxies. Among various
emissions at different wavelength bands, the [O {\sc
iii}]$\lambda$5007, infrared and hard X-ray (2-10kev) continua
have been found (Dahari \& De Robertis  1988, Keel et al.  1994,
Mulchaey et al.  1994, Alonso-Herrero et al.  1997) to indicate
similar distributions for Sy1s and Sy2s, implying their isotropic
properties so as to be good indicators of the intrinsic nuclear
luminosity.

Recent investigation have found that about  (30-50)\% of Seyfert 2
galaxies show nuclear starburst activities (Storchi-Bergmann et
al. 2000; Gonzalez Delgado et al. 2001).It follows that the
infrared radiation may not be a good indicator for AGN luminosity,
because of the contamination from nuclear star-forming activities.
On the other hand,  the incomplete isotropy of {\rm [O {\sc iii}]}
emission caused by obscuration from host galaxies brings about a
suggestion (Maiolino et al.  1998)  to take the
extinction-corrected {\rm [O {\sc iii}]} luminosity as an
indicator of the nuclear activity.

The same is true for  the hard X-ray (2-10kev) continua, in a
sense that they are heavily absorbed. About half of Sy2s in the
Local Universe are Compton-thick with absorbing column densities
N$_{\rm H} > 10^{24}$ cm$^{-2}$ (Maiolino et al. 1998; Bassani et
al. 1999; Risaliti, Maiolino \& Salvati 1999). In the case of
Compton-thick Seyfert 2 galaxies, the  direct hard X-ray continua
are  completely absorbed, implying that this indicator of AGN
activities works well for Compton-thin sources only. What is more,
the effect of absorption on X-ray emission should be corrected,
such as deriving the extinction-corrected {\rm [O {\sc iii}]}
mentioned above.

\subsection{Statistics on hard X-ray $-$ radio emissions}

Following the above considerations, we perform statistics on
various indicators of AGN activities for SPSS1 and SPSS0
subsamples. The results are illustrated in Fig 3 - Fig 6, where
one could find cumulative distributions of extinction-corrected
{\rm [O {\sc iii}]} luminosities (in Fig 3),  mid-infrared flux
ratios (f$_{60}$/f$_{25}$), 25 $\mu$m luminosities and radio
powers at 1.49 GHz (in Fig 4a, b and c), respectively. It is
evident from the comparisons that SPSS1 differ from SPSS0
dramatically. The Sy2s {\it with} PBL (SPSS1) show more powerful
{\rm [O {\sc iii}]} emission, much hotter mid-infrared color and
significant excesses of 25 $\mu$m and radio emissions. Indeed, the
KS test or GGW test shows that the SPSS1 and SPSS0 subsamples are
drawn from the same parent population with a possibility P$_{\rm
null} < 0.0001$. The statistical results are given in Table 3.

In Fig 5, we show a plot of the absorption-corrected 2-10 kev
luminosity vs extinction-corrected {\rm [O {\sc iii}]} luminosity
for Compton-thin Seyfert 2 galaxies in SPSS1 (in filled circles)
and SPSS0 (in open circles) subsamples. It is instructive to find
that Compton-thin Sy2s {\it with} and {\it without} PBL are
located in separate regions. The mean $\log L_{\rm 2-10kev}$ for
SPSS1 and SPSS0 sources are 43.0 $\pm$ 0.14 erg s$^{-1}$ and 41.22
$\pm$ 0.32 erg s$^{-1}$, respectively, clearly indicating that the
nuclear activities in SPSS1 are much more powerful than that in
SPSS0. The GGW test of P$_{\rm null} <0.0001$ provides full
support to this claim.

Probably related to the above finding is what we show in Fig 6, a
plot of extinction-corrected {\rm [O {\sc iii}]} emission against
Eddington luminosity (L$_{\rm Edd}$) for SPSS1 and SPSS0
subsamples. In accordance with the strong tendency manifested in
Fig 5, what Fig 6 illustrates is that the majority of SPSS1
sources have higher L$_{\rm [O III]}$/L$_{\rm Edd}$. That is, 11
out of 14 SPSS1 sources have L$_{\rm [O III]}$/L$_{\rm Edd}
> 10^{-4}$, while 12 Of 17 SPSS0 $< 10^{-4}$. It suggests that the
accretion rates in SPSS1 sources might be higher than that in
SPSS0 on average.

\section{Discussions}

It has long been known that L$_{\rm FIR}$/L$_{\rm B}$ is a good
star formation indicator for spiral galaxies (see, e.g. Gu et al.
1999; Lei et al. 2000). We find that the distributions of L$_{\rm
FIR}$/L$_{\rm B}$ for SPSS1 and SPSS0 are similar, with the mean
L$_{\rm FIR}$/L$_{\rm B}$ of -0.275 and -0.298, respectively. It
implies that the detectability of PBL does not directly relate to
the star-forming activities in  host galaxies. In fact, we may
list  both SPSS1 and SPSS0 sources showing nuclear starburst
activities, e.g. Mrk 477, Mrk 533, IC 3639 in SPSS1, and NGC 5135,
NGC 7130 in SPSS0 (Gonzalez Delgado et al. 2001).

The detectability of a hidden BLR in Sy2s has been suggested
(Heisler et al. 1997) having relation to the inclination of the
torus determined by the 60 $\mu$m to 25 $\mu$m flux ratio.
However,  by comparing the column densities inferred from hard
X-rays, Alexander (2001) argued against the direct relation of the
detectability of the PBL to IR colors and claimed that the
contribution from host galaxies  would both make the IR color
cooler and dilute the nuclear activity. In the mean time, Gu,
Maiolino \& Dultzin-Hacyan (2001) find that the detectability of
PBL depends on the relative contribution/dilution from  host
galaxies.  Lumsden et al. (2001) also suggested that the
detectability of PBL is largely due to a combination of AGN
activity, the obscuring density and the relative contribution of
host galaxies.

By analyzing various indicators of AGN activities, we find in this
work that mid-infrared, radio, hard X-ray, and [O {\sc iii}]
luminosities of SPSS1 are significantly larger than  SPSS0,
indicating obviously that the detectability of PBL is most likely
due to the AGN activities. Indeed, the mean absorption-corrected
X-ray luminosity in  2-10kev energy band for SPSS1 ,  L$_{\rm
2-10kev}^{c} \sim$ 10$^{43}$ erg s$^{-1}$, is the typical X-ray
luminosity for Seyfert 1 galaxies  (Nandra et al., 1997). The mean
L$_{\rm 2-10kev}^{c}$ of SPSS0, on the contrary, is nearly 2 order
of magnitudes smaller, most of them show very low accretion rates
as Fig 6 illustrated.  What we can infer from these analyses is
that the majority of SPSS0 (Sy2s {\it without} PBL) sources are
probably low-luminosity AGNs (LLAGNs). The more powerful the AGNs
are, the easier it is to detect PBL, which is also consistent with
our finding of the morphological difference of Sa from Sab between
SPSS1 and SPSS0. Galaxies with earlier Hubble types have mirrors
that are easier to be detected.

Our sample is an amalgamation of different observations with
diverse quality of spectropolarimetric data,  varying from object
to object determined by the  brightness, observers, integration
time, and a host of other factors. For example, Ruiz et al. (1994)
have found no evidence for broad H$_\alpha$ and H$_\beta$
components in the polarized light for Mrk~334, yielding a maximum
of $\sim$ 0.6\% of the light in H$_\alpha$ as polarized broad-line
flux. For other Sy2s with detected polarized broad lines, the mean
percentage of 1\% - 5\% has been reported. Needless to say that it
is necessary to evaluate the sensitivities to detect  polarized
broad lines in different groups from which our data are taken. The
difficult situation we are faced with is that only one group
(Young et al 1996) has presented detailed information for their
polarized broad emission components, so that we can derive their
sensitivity $-$ the detected minimum polarized broad H$_\alpha$
flux is 1.2 10$^{-15}$ erg s$^{-1}$ cm$^{-2}$ for NGC 7674 and
their mean percentage of polarized broad H$_\alpha$ emission to
total H$_\alpha$ emission is 4.7 $\pm$ 0.8 \%. For three groups
(Moran et al 2000, Moran et al 2001, Tran 2001, Lumsden et al
2001) reporting no detailed polarized properties, to compare the
common sources that they observed would be an adequate approach to
evaluate their sensitivities. The analyzed results are summarized
in Table 4.

At first glance, one could find that there are 16 common Sy2s in
both Tran's and Lumsden's samples and 13 in both Tran's and
Moran's samples, and that detectabilities of PBL for these common
sources are the same, which indicates comparable sensitivities of
these three samples. On the other hand, 3 Sy2s (IRAS 00521-7054,
IRAS 04385-0828 and NGC 5506) with PBL reported by  Tran (2001)
are not listed  in Young's sample as Sy2s {\it with} PBL.  What is
more, two Sy2s (NGC 5347 and NGC 5929) listed as "non-HBLR" Sy2s
in Tran's sample have been discovered (Moran et al.  2001) showing
faint polarized broad emission lines with high S/N data.

The conclusions we could reach from the above analyses is that the
sensitivity of Moran's sample (2001) is higher than that of Tran's
(2001), both of which are higher than that of Young's sample
(1996). Since the polarized fluxes of broad emission lines
represent a certain fraction of the total flux, typically a few
percents. it should be true to say that the Sy2s {\it without}
detection of PBL  would be preferentially harder to be detected
for a given sensitivity. In other words, for less luminous Sy2s
without detection of PBL, one would need to reach sensitivities
that would allow us to detect the {\it same percentage
polarization}. Therefore, to attribute those Sy2s without
detection of PBL to non-hidden BLR in those sources would be
premature. What we can point out clearly is that {\it we have
found no compelling evidence for the presence of two types of Sy2s
as claimed by Tran(2001)}.

Antonucci (2001) has repeatedly made the point that, to be
compelling, evaluations of one sample vs. another must be made for
samples that are selected by a property which is thought to be
isotropic. Due to the limitation of our sample selection
(heterogeneous with different spectropolarimetric sensitivities),
it is impossible to interpret our statistical results in the
configuration of testing the unified model for AGN. On the
contrary, the conclusions we reached above force us to claim at
the present time that {\it we have found no compelling evidence to
argue for modifications to the unified model of AGN}. The
non-detection of PBL is probably due to the low AGN activities.
Based on the analyses presented in this paper,  we can not rule
out the possibility of the combination of the weakness of AGN and
commensurate obscuration/dilution from host galaxies (Lumsden \&
Alexander 2001, Gu Maiolino \& Dultzin-Hacyan  2001).

\section{Conclusions}
In this paper, we collect radio, infrared, optical and hard X-ray
data for a sample of 90 Seyfert 2 galaxies with
spectropolarimetric observations. Out of these 90 objects, 41 show
polarized broad lines (most likely ascribed to scattering of the
broad line region) and 49 do not. Compared to Sy2s without PBL,
Sy2s with PBL tend to be earlier-type spirals, and show warmer
mid-infrared color and significant excess of emissions (including
the hard X-ray(2-10kev), [O {\sc iii}]$\lambda$5007, infrared (25
$\mu$m) and radio), while their distributions of blue luminosity
and absorbing column density are similar. Our analyses suggest
that the majority of Sy2s without PBL are those sources having
less powerful AGN activities, most likely caused by low accretion
rate. It implies that the detectability of the polarized broad
emission lines in Sy2s may depend on their central AGN activities
in most cases. Based on the available data, we find no compelling
evidence for the presence of two types of Sy2s, one type of them
has been proposed to be intrinsically different from Sy2s claimed
in Unification Model.

\acknowledgements  The authors are very grateful to the anonymous
referee for his$/$her critical comments and instructive
suggestions, which significantly strengthened the analyses in this
paper. We also thank Luis Ho, Xiaoyang Xia and Zugan Deng for
their valuable discussion and suggestion. This work has been
supported by the National Natural Science Foundation of China
under grant 10103001 and the National Key Basic Research Science
Foundation (NKBRSF G19990754). This research has made use of
NASA's Astrophysics Data System Abstract Service and the NASA/IPAC
Extragalactic Database (NED) which is operated by the Jet
Propulsion Laboratory, California Institute of Technology, under
contract with the National Aeronautics and Space Administration.

\newpage

\begin{figure}
\plotone{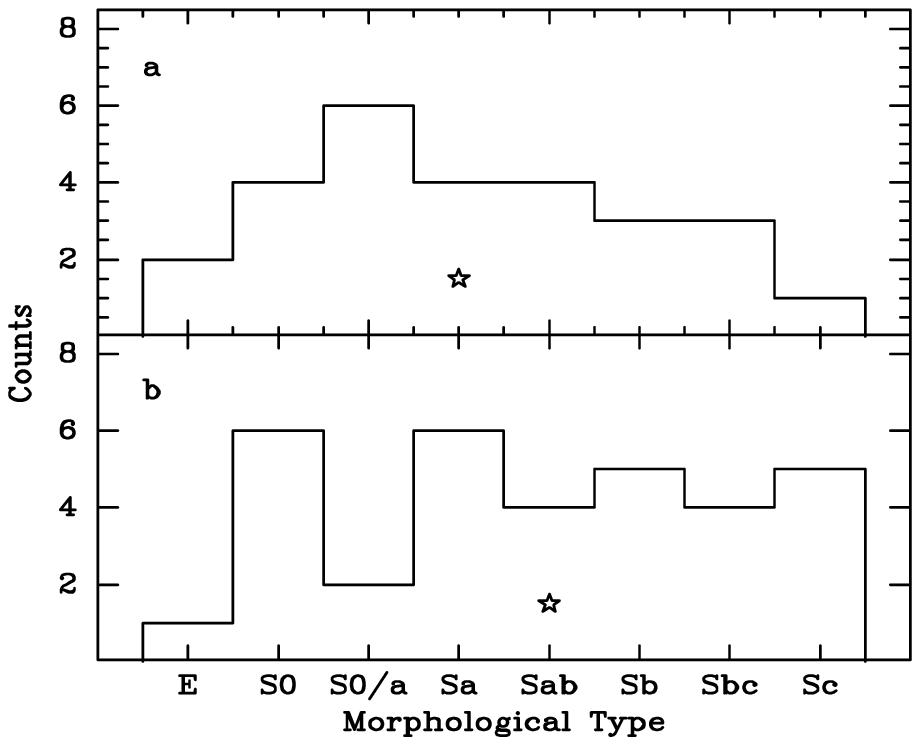} \caption{The histogram distribution of the
morphological types for Sy2s with PBL (a) and for Sy2s without PBL
(b). \label{fig1}}
\end{figure}

\clearpage

\begin{figure}
\plotone{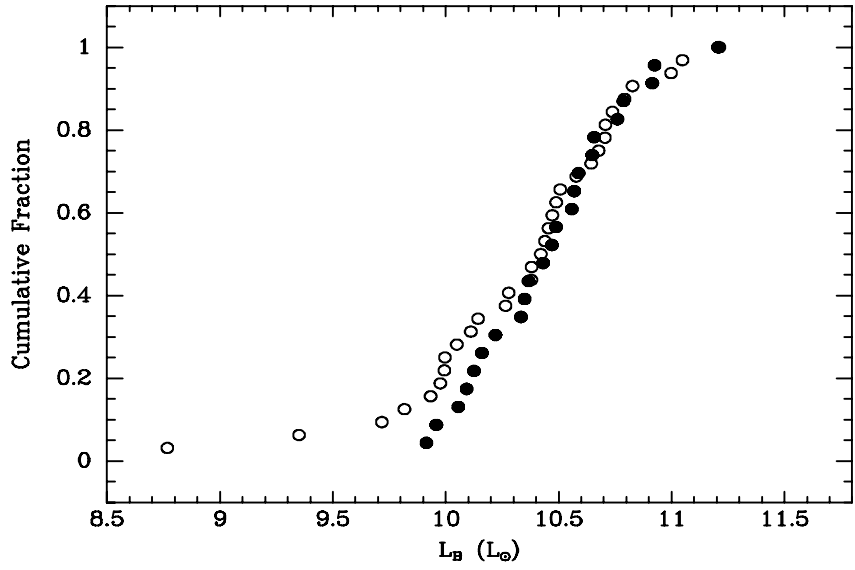} \caption{The cumulative distribution of the
blue luminosities for Sy2s with PBL (filled circles) and for Sy2s
without PBL (open circles ). \label{fig2}}
\end{figure}

\clearpage

\begin{figure}
\plotone{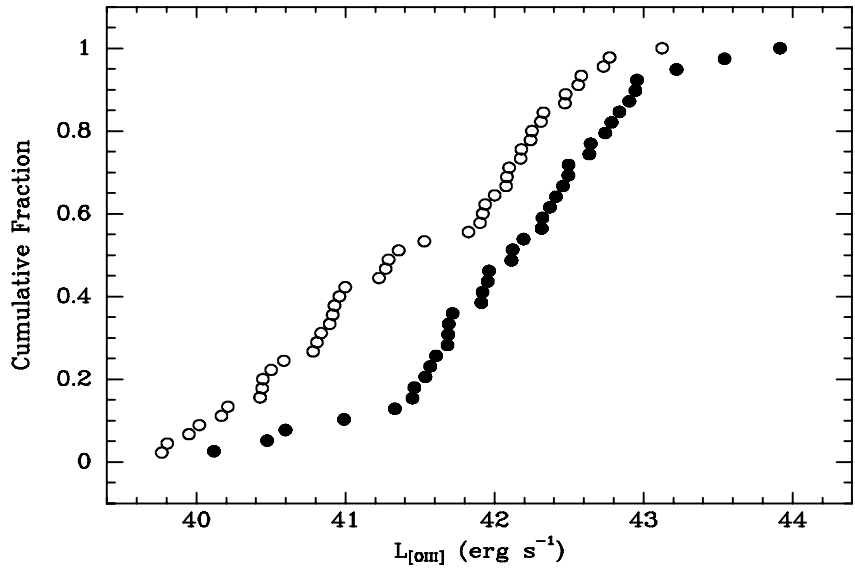} \caption{The cumulative distribution of
luminosities of extinction-corrected [O {\sc iii}]$\lambda$5007
emission. Symbols have the same coding as in Fig 2. \label{fig3}}
\end{figure}

\clearpage

\begin{figure}
\plotone{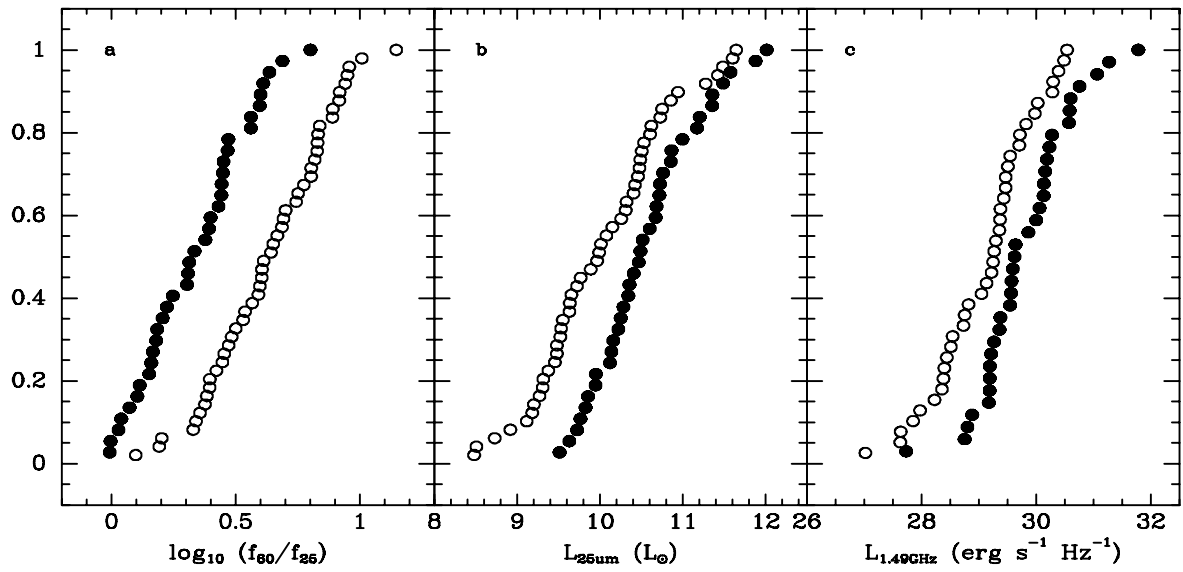} \caption{The cumulative distribution of
infrared ratio f$_{60}$/f$_{25}$(a), mid-infrared (25 $\mu$m)
luminosity (b) and radio 1.49 GHz continuum power (c). Symbols
have the same coding as in Fig 2. \label{fig4}}
\end{figure}

\clearpage

\begin{figure}
\plotone{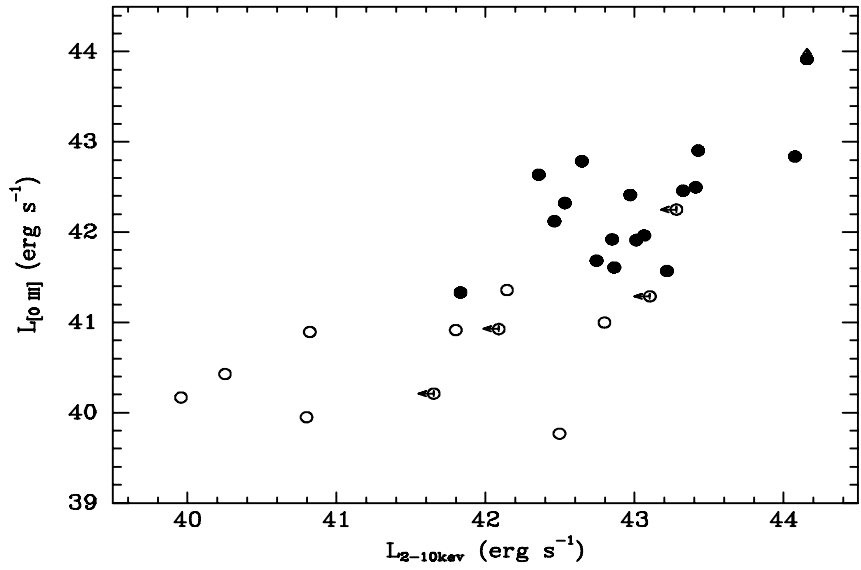} \caption{The distribution of
absorption-corrected 2-10kev luminosities against
extinction-corrected [O {\sc iii}] luminosities for Compton-thin
Seyfert 2 galaxies. Symbols have the same coding as in Fig 2.
\label{fig5}}
\end{figure}

\clearpage

\begin{figure}
\plotone{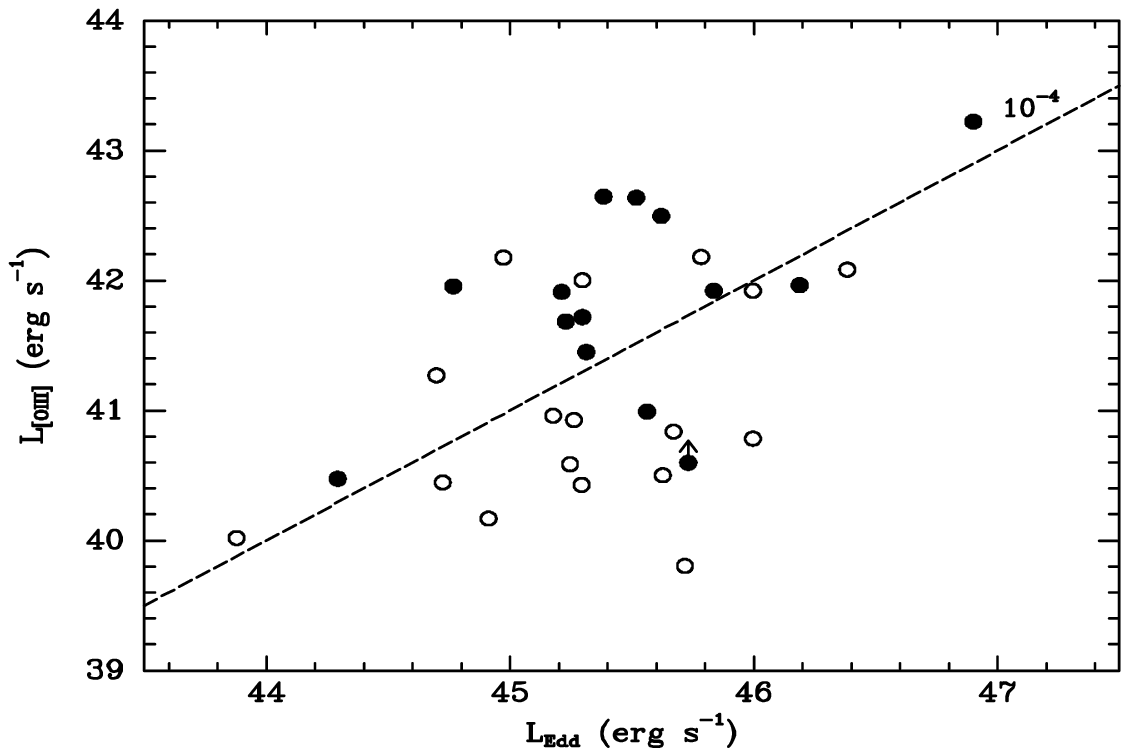} \caption{The distribution of Eddington
luminosities against [O {\sc iii}] luminosities. Symbols have the
same coding as in Fig 2. \label{fig6}}
\end{figure}

\clearpage

\begin{deluxetable}{lcccccccc}
  \tabletypesize{\scriptsize}
  \tablecaption{Infrared and radio data for Seyfert 2 galaxies}
  \tablewidth{0pt}
  \tablehead{\colhead{Name} &
 \colhead{f$_{\rm 60}$/f$_{\rm 25}$} &
  \colhead{L$_{\rm 25\mu m}$$^{\rm a}$} & \colhead{L$_{\rm FIR}$$^{a}$} &
  \colhead{L$_{\rm 1.49GHz}$$^{b}$}     & \colhead{Type}     & \colhead{PBL?}  & \colhead{References} \\
  \hspace*{10.mm}(1)&(2)&(3)&(4)&(5)&(6)&(7)&(8) }
  \startdata
    Mrk  334     &      4.038 &   10.505 &   10.674 &   29.434  & pec            &no& R94 \\
    NGC   34     &      6.756 &   10.751 &   11.143 &   29.705  & Sc             &no& H97  \\
IRAS 00198-7926  &       2.488 &   11.603 &   11.547 &           & pec            &no& T01  \\
    Mrk  348     &        1.419 &   10.161 &    9.855 &   30.132  & SA(s)0/a       &yes& M90  \\
IRAS 00521-7054  &        1.270 &   11.357 &   11.064 &           & E-S0           &yes& T01  \\
    NGC  424     &        1.183 &   10.122 &    9.740 &   28.752  & (R)SB(r)0/a    &yes& M00  \\
    NGC  513     &        6.326 &    9.856 &   10.348 &   29.621  & Sbc           &yes& T01  \\
    NGC  591     &        4.327 &    9.827 &   10.094 &   29.187  & (R')SB0/a      &yes& M00  \\
    Mrk  573     &        1.593 &   10.150 &    9.906 &   29.133  & (R)SAB(rs)0    &no& T01  \\
IRAS 01475-0740  &         1.095 &   10.252 &    9.798 &   30.278  & E-S0           &yes& T01  \\
    NGC  788     &        0.992 &    9.765 &    9.336 &           & SA(s)0/a       &yes& K98  \\
    NGC 1068     &        2.160 &   10.865 &   10.789 &   30.130  & (R)SA(rs)b     &yes& A85  \\
    NGC 1144     &        8.724 &   10.459 &   11.087 &   30.391  & S pec          &no& T01  \\
    Mrk 1066     &        4.857 &   10.321 &   10.578 &   29.467  & (R)SB(s)0      &no& M00  \\
IRAS 02581-1136  &        1.067 &   10.407 &   10.168 &   29.175  & SAB(rs)a       &yes& T01  \\
    NGC 1241     &        8.294 &    9.767 &   10.381 &           & SB(rs)b        &no& T01  \\
    NGC 1320     &        2.189 &    9.722 &    9.644 &   27.974  & Sa             &no& T01  \\
    NGC 1358     &        3.048 &    9.115 &    9.302 &   28.820  & SAB(r)0/a      &no& M00  \\
    NGC 1386     &        4.027 &    8.729 &    8.962 &   27.637  & SB(s)0         &no& M00  \\
IRAS 03362-1642  &         2.133 &   10.606 &   10.588 &   29.370  &                &no& T01  \\
IRAS 04103-2838  &        3.388 &   11.646 &   11.724 &   30.539  &                &no& Y96  \\
IRAS 04210+0400  &         2.378 &   10.487 &   10.697 &   30.295  &                &no& Y96  \\
IRAS 04229-2528  &        3.687 &   10.478 &   10.633 &   29.547  &                &no& Y96  \\
IRAS 04259-0440  &        2.798 &   10.311 &   10.297 &           & E-S0           &no& H97  \\
IRAS 04385-0828  &         1.766 &   10.357 &   10.125 &   28.882  & S0             &yes& T01  \\
    NGC 1667     &       8.914 &    9.966 &   10.636 &   29.506  & SAB(r)c        &no& M00  \\
    NGC 1685     &        4.477 &    9.456 &    9.726 &   28.753  & SB(r)0/a       &no& M00  \\
IRAS 05189-2524  &       3.986 &   11.578 &   11.721 &   29.999  & pec            &yes& H97  \\
    Mrk    3     &       1.300 &   10.515 &   10.171 &   30.600  & S0             &yes& M90  \\
    NGC 2273     &        4.881 &    9.631 &    9.932 &   28.803  & SB(r)a         &yes& M00  \\
    ESO 428-G014 &        2.486 &    9.376 &    9.370 &   28.542  & SA0            &no& M00  \\
    Mrk 1210     &        0.984 &   10.283 &    9.792 &   29.571  & Sa             &yes& T92  \\
IRAS 08277-0242  &        3.449 &   10.620 &   10.742 &   30.028  &                &no& Y96  \\
    NGC 3081     &              &          &          &   27.731  & (R1)SAB(r)0/a &yes& M00  \\
    NGC 3079     &       14.014 &    9.521 &   10.336 &   29.359  & SB(s)c         &no& T01  \\
    NGC 3281     &        2.645 &   10.257 &   10.250 &   29.248  & SAB(rs+)a      &no& M00  \\
IRAS 10340+0609  &        1.556 &    9.309 &    9.240 &           &                &no& M00  \\
    NGC 3362     &             &          &          &   29.376  & SABc           &no& T01  \\
    UGC 6100     &        2.842 &   10.012 &   10.182 &   29.265  & Sa             &no& T01  \\
IRAS 11058-1131  &        2.387 &   10.763 &   10.700 &   29.592  &                &yes& V01  \\
    NGC 3660     &        7.780 &    9.315 &    9.905 &   28.513  & SB(r)bc        &no& T01  \\
    NGC 3982     &         8.258 &    8.918 &    9.531 &   28.226  & SAB(r)b        &no& M00  \\
    NGC 4117     &               &          &          &   27.020  & S0             &no& M00  \\
    Was  49b     &        1.446 &   10.856 &   10.617 &   30.755  &                &yes& T92  \\
    NGC 4388     &        2.939 &   10.139 &   10.244 &   29.188  & SA(s)b         &yes& H97  \\
    NGC 4501     &       5.953 &    9.990 &   10.559 &   29.468  & SA(rs)b        &no& T01  \\
    NGC 4507     &       3.953 &    9.952 &   10.168 &   29.190  & SAB(s)ab       &yes& M00  \\
    IC  3639     &        2.953 &   10.222 &   10.308 &   29.265  & SB(rs)bc       &yes& H97  \\
    NGC 4941     &       4.098 &    8.480 &    8.812 &   27.629  & (R)SAB(r)ab    &no& M00  \\
    MCG -3-34-64 &       2.058 &   10.683 &   10.549 &   30.158  & SB?            &yes& T01  \\
    NGC 5135     &       6.732 &   10.422 &   10.886 &   29.824  & SB(l)ab        &no& H97  \\
    NGC 5194     &       6.587 &    9.480 &   10.035 &   28.441  & SA(s)bc        &no& T01  \\
    NGC 5252     &             &          &          &   29.212  & L              &yes& Y96  \\
    NGC 5256     &        6.363 &   10.729 &   11.138 &   30.283  & pec            &no& T01  \\
    NGC 5283     &              &          &          &   28.355  & S0             &no& M00  \\
    Mrk 1361     &       3.905 &   10.406 &   10.569 &   29.297  & SB             &no& H97  \\
IRAS 13452-4155  &       2.272 &   10.857 &   10.735 &           &                &no& Y96  \\
    NGC 5347     &         1.253 &    9.656 &    9.442 &   27.852  & (R')SB(rs)ab   &yes& M01  \\
    Mrk  463E    &        1.514 &   11.356 &   11.087 &   31.272  &                &yes& M90  \\
    Circinus     &       3.640 &    9.513 &    9.662 &           & SA(s)b         &yes& V01  \\
    NGC 5506     &        2.014 &    9.947 &    9.809 &   29.360  & Sa             &yes& T01  \\
    NGC 5643     &        5.025 &    9.552 &    9.895 &           & SAB(rs)c       &no& M00  \\
    NGC 5695     &       4.388 &    9.203 &    9.606 &   28.397  & SBb            &no& M00  \\
    Mrk  477     &        2.507 &   10.674 &   10.672 &   30.230  &                &yes& T92  \\
    NGC 5728     &       10.195 &    9.636 &   10.275 &   29.047  & (R1)SAB(r)a   &no& M00  \\
    ESO 273-IG04 &       2.767 &   11.172 &   11.174 &           &                &yes& V01  \\
    NGC 5929     &      5.642 &    9.889 &   10.253 &   29.220  & Sab            &yes& M01  \\
    NGC 5995     &        2.824 &   10.724 &   10.808 &   29.561  & S(B)c          &yes& T01  \\
IRAS 15480-0344  &         1.468 &   10.601 &   10.574 &   29.863  & S0             &yes& T01  \\
IRAS 17345+1124  &         2.469 &   11.487 &   11.855 &   31.782  &                &yes& V01  \\
    NGC 6552     &        3.644 &   10.342 &   10.491 &   29.640  & SB?            &yes& T01  \\
IRAS 19254-7245  &        3.955 &   11.484 &   11.667 &           &                &no& H97  \\
    NGC 6890     &        4.923 &    9.480 &    9.873 &           & (R')SA(r)ab   &no& M00  \\
IRAS 20210+1121  &        2.421 &   11.424 &   11.337 &   30.485  &                &no& Y96  \\
IRAS 20460+1925  &        1.672 &   12.014 &   11.865 &   31.064  &                &yes& Y96  \\
    IC  5063     &        1.528 &   10.468 &   10.170 &           & SA(s)0+        &yes& I93  \\
    NGC 7130     &        7.757 &   10.532 &   11.038 &   29.977  & Sa             &no& H97  \\
    NGC 7172     &        6.372 &    9.632 &   10.100 &   28.731  & Sa             &no& H97  \\
IRAS 22017+0319  &        1.607 &   11.206 &   10.953 &   30.060  &                &yes& T01  \\
    NGC 7212     &        4.083 &   10.489 &   10.736 &   30.188  &                &yes& T92  \\
    MCG -3-58-7  &        2.774 &   10.718 &   10.773 &   29.374  & (R')SAB(s)0/a  &yes& T01  \\
IRAS 23060+0505  &        2.700 &   11.881 &   11.833 &   30.584  & pec            &yes& V01  \\
    NGC 7496     &        5.536 &    9.531 &    9.890 &           & (R':)SB(rs)bc  &no& H97  \\
IRAS 23128-5919  &        6.792 &   11.276 &   11.666 &           &                &no& Y96  \\
    IC  5298     &        4.654 &   10.943 &   11.205 &   29.715  &                &no& L01  \\
    NGC 7582     &       6.895 &   10.078 &   10.531 &           & (R'1)SB(s)ab  &no& H97  \\
    NGC 7590     &        9.065 &    9.182 &    9.844 &           & S(r?)bc        &no& H97  \\
    NGC 7672     &       2.961 &    9.271 &    9.646 &   28.382  & Sb             &no& M00  \\
    NGC 7674     &       2.809 &   10.996 &   11.057 &   30.572  & SA(r)bc        &yes& M90  \\
    NGC 7682     &        2.034 &    9.725 &    9.678 &   29.545  & SB(r)ab        &yes& T01  \\
  \enddata

\tablenotetext{a}{Luminosities in units of L$_\odot$.}
\tablenotetext{b}{Radio 1.49GHz luminosity in unit of erg s$^{-1}$
Hz$^{-1}$} \tablecomments{{\sc References for PBL}: A85 =
Antonucci \& Miller (1985); H97 = Heisler et al. (1997); I93 =
Inglis et al. (1993); K98 = Kay \& Moran (1998); L01 = Lumsden et
al. (2001); M90 = Miller \& Goodrich (1990); M00 = Moran et al.
(2000); M01 = Moran et al. (2001); R94 = Ruiz et al. (1994); T92 =
Tran et al. (1992); T01 = Tran (2001); V01 = Veron \& Veron
(2001); Y96 = Young et al. (1996).}
\end{deluxetable}

\clearpage

\begin{deluxetable}{lcccccccc}
  \tabletypesize{\scriptsize}
  \tablecaption{Optical and hard X-ray data for Seyfert 2 galaxies}
  \tablewidth{0pt}
  \tablehead{
  \colhead{Name}  &\colhead{L$_{\rm B}$$^{\rm a}$}  &  \colhead{L$_{\rm H_{\beta}}$$^{\rm b}$ }        &
  \colhead{L$_{\rm [O III]}$$^{\rm c}$}&  \colhead{References}                         &
  \colhead{log$_{10}$ N$_{\rm H}$}            &  \colhead{L$_{\rm 2-10kev}$$^{\rm d}$} &
  \colhead{References}                  &  \colhead{L$_{\rm Edd}$$^{\rm e}$}\\
      \hspace*{10.mm}(1) & (2) & (3) &(4) & (5) &(6) & (7)& (8) &
      (9)
  }
  \startdata
    Mrk  334    &  &  40.507&      41.289&d88&    20.643& $<$  43.102&p02&          \\
    NGC   34    &  &  39.056&      42.772&d88&          &          &   &          \\
IRAS 00198-7926 &  &  40.532&      42.562&d92& $>$ 24.0&          &r00&          \\
    Mrk  348    & 10.220  & 39.870&      41.912& d88 &   23.041&    43.011&b99 &  45.211 \\
IRAS 00521-7054 &   & 40.940&      42.743& d92 &         &          &    &         \\
    NGC  424    & 10.160 &  40.350&      41.462& m94 &   24.176&          &c00 &         \\
    NGC  513    & 10.760 &  40.327& $>$  40.597& c94 &         &          &    &  45.730 \\
    NGC  591    & 10.349 &  39.972&      41.954& d88 &         &          &    &  44.767 \\
    Mrk  573    & 10.455 &  40.452&      42.001&d88&          &          &   &   45.296 \\
IRAS 01475-0740 &  &  39.784&      41.689& d92 &         &          &    &         \\
    NGC  788    & 10.569 &  39.735&      40.990& w92 &         &          &    &  45.562 \\
    NGC 1068    & 10.787  & 40.675&      42.645& m94 &$>$ 25.0&          &b99 &  45.383 \\
    NGC 1144    & 11.048 &  39.720&      42.251&d88&    20.699& $<$  43.281&p02&          \\
    Mrk 1066    & 10.379 &  40.205&      42.175&m94& $>$  24.0&          &r99&   44.972 \\
IRAS 02581-1136 &  &  40.307&      41.536& d92 &         &          &    &         \\
    NGC 1241    & 10.827&$<$  38.529&    42.472&d85&          &          &   &          \\
    NGC 1320    & &   39.346&      40.959&m94&          &          &   &   45.176 \\
    NGC 1358    & 10.578&   39.757&      40.783&m94&          &          &   &   45.996 \\
    NGC 1386    & 9.350 &  38.787&      40.586&m94& $>$  25.0&          &b99&   45.246 \\
IRAS 03362-1642 &   & 39.878&      41.529&d92&          &          &   &          \\
IRAS 04103-2838 &   &       &          &   &          &          &   &          \\
IRAS 04210+0400 &   & 40.463&      42.328&y96&          &          &   &          \\
IRAS 04229-2528 &   & 38.861&      41.903&y96&          &          &   &          \\
IRAS 04259-0440 &   & 38.908&      40.440&v95&          &          &   &          \\
IRAS 04385-0828 & &$<$  39.102&    40.117& d92 &         &          &    &         \\
    NGC 1667    & 10.792 &  39.889&      41.921&s95& $>$  24.0&          &b99&   45.996 \\
    NGC 1685    &  &  40.081&      42.582&c94&          &          &   &          \\
IRAS 05189-2524 &  &  39.633&      42.459& v95 &   22.690&    43.325&b99 &         \\
    Mrk    3    & 10.333  & 41.018&      43.221& m94 &   24.041&          &b99 &  46.900 \\
    NGC 2273    & 10.125  & 40.314&      41.449& l92 &$>$  25.0&          &b99 &  45.313 \\
  ESO 428-G014  & 9.976  & 39.621&      41.937&a91& $>$  25.0&          &m98&          \\
    Mrk 1210    &  9.958 & 40.472&      42.195& t91 &$>$  24.0&          &b99 &         \\
IRAS 08277-0242 &   &       &          &   &          &          &   &          \\
    NGC 3081    &  10.092 & 39.908&      41.331& m94 &   23.820&    41.831&b99 &         \\
    NGC 3079    &  10.439 & 37.145&      40.427&h97&    22.204&    40.251&b99&   45.293 \\
    NGC 3281    &  10.707 & 38.044&      40.998&m94&    23.903&    42.798&b99&          \\
IRAS 10340+0609 &   &       &          &   &          &          &   &          \\
    NGC 3362    &  10.998 & 40.001&      41.269&p96&          &          &   &   44.697 \\
    UGC 6100    &  10.738 & 40.592&      42.180&c94&          &          &   &   45.784 \\
IRAS 11058-1131 &   & 41.199&      42.316& d92 &$>$  24.0&          &r00 &         \\
    NGC 3660    &   & 39.367&      40.914&s95&    20.255&    41.800&tar&          \\
    NGC 3982    & 9.933  & 38.174&      40.019&p96&          &          &   &   43.878 \\
    NGC 4117    & 8.767  &       &          &   &          &          &   &   44.767 \\
    Was  49b    &   & 41.257&      42.412& m92 &   22.799&    42.970&a00 &         \\
    NGC 4388    & 10.925  & 39.691&      41.684& m94 &   23.623&    42.744&b99 &  45.228 \\
    NGC 4501    & 11.212  & 38.873&      39.804&h97&          &          &   &   45.717 \\
    NGC 4507    & 10.588  & 40.430&      41.569& m94 &   23.462&    43.217&b99 &         \\
    IC  3639    & 10.367  & 40.255&      42.113& s95 &$>$  25.0&          &r99 &         \\
    NGC 4941    & 9.817  & 38.836&      40.894&s89&    23.653&    40.820&b99&          \\
    MCG -3-34-64&   & 41.261&      42.322& d88 &   23.881&    42.531&b99 &         \\
    NGC 5135    &  10.705 & 40.153&      42.311&m94& $>$  24.0&          &b99&          \\
    NGC 5194    &  10.472 & 37.919&      40.168&h97&    23.699&    39.956&b99&   44.911 \\
    NGC 5252    &  10.558 & 39.924&      41.963& y96 &   22.633&    43.064&b99 &  46.188 \\
    NGC 5256    &   & 40.295&      41.825&d88& $>$  24.0&          &b99&          \\
    NGC 5283    &  9.717 & 39.596&      40.836&d88&          &          &   &   45.670 \\
    Mrk 1361    &  10.049 & 40.285&      42.242&v95&          &          &   &          \\
IRAS 13452-4155 &   & 40.475&      42.098&y96&          &          &   &          \\
    NGC 5347    & 9.994&$<$  38.803&    40.446&d92& $>$  24.0&          &r99&   44.723 \\
    Mrk  463E   & 11.206 &  41.223&      42.785& d88 &   23.204&    42.646&b99 &         \\
    Circinus    & 9.914 &  39.402&      40.474& o94 &$>$  24.0&          &b99 &  44.293 \\
    NGC 5506    & 10.056 &  39.585&      41.608& s95 &   22.531&    42.863&b99 &         \\
    NGC 5643    & 10.421 &  39.063&      41.225&m94& $>$  25.0&          &b99&          \\
    NGC 5695    & 10.380 &  39.383&      40.501&d88&          &          &   &   45.626 \\
    Mrk  477    & 10.488 &  41.626&      43.543& d88 &$>$  24.0&          &b99 &         \\
    NGC 5728    & 10.644 &  39.946&      42.083&s95&          &          &   &   46.383 \\
    ESO 273-IG04&  &  41.129&      42.372& y96 &         &          &    &         \\
    NGC 5929    &  &  39.545&      40.926&m94&    20.763& $<$  42.088&p02&   45.262 \\
    NGC 5995    &  &  42.104&      42.904& l01 &   21.934&    43.426&l01 &         \\
IRAS 15480-0344 &  &  40.021&      42.946& d92 &         &          &    &         \\
IRAS 17345+1124 &  &  41.961&      42.956& t99 &         &          &    &         \\
    NGC 6552    & 10.649 &        &      42.122& b99 &   23.778&    42.461&b99 &         \\
IRAS 19254-7245 &  &  40.044&      42.731&v02& $>$  24.0&          &r00&          \\
    NGC 6890    & 10.111 &  39.059&      40.808&m94&          &          &   &          \\
IRAS 20210+1121 &  &  41.454&      43.124&y96& $>$  25.0&          &b99&          \\
IRAS 20460+1925 &  &  41.113&      42.838& y96 &   22.398&    44.076&b99 &         \\
    IC  5063    & 10.470 &  40.322&      41.919& m94 &   23.380&    42.849&b99 &  45.836 \\
    NGC 7130    & 10.677 &  40.547&      42.477&s95& $>$  24.0&          &r00&          \\
    NGC 7172    & 10.279 &  38.733&      39.767&v97&    22.934&    42.495&b99&          \\
IRAS 22017+0319 &  &  41.225&      42.497& d92 &   21.301&    43.410&r00 &         \\
    NGC 7212    & 10.657 &  40.941&      42.636& m94 &   23.653&    42.356&a00 &  45.517 \\
    MCG -3-58-7 & &$<$  40.484&    41.692& d92 &         &          &    &         \\
IRAS 23060+0505 & &$<$  41.005& $>$43.916& d92 &   22.924&    44.155&    &         \\
    NGC 7496    & 10.143&   39.382&      40.211&y96&    22.699& $<$  41.652&p96&          \\
IRAS 23128-5919 & &         &          &   &          &          &   &           \\
    IC  5298    & 10.489&   39.927&      42.078&v95&          &          &   &          \\
    NGC 7582    & 10.507&   39.912&      41.357&s95&    23.079&    42.144&b99&          \\
    NGC 7590    & 10.265&   39.152&      39.949&s95& $<$  20.964&    40.798&b99&          \\
    NGC 7672    & 9.996&         &          &   &          &          &   &   44.831 \\
    NGC 7674    & 10.914&   40.808&      42.495& d88 &$>$  25.0&          &b99 &  45.620 \\
    NGC 7682    & 10.431&   40.197&      41.718& c94 &         &          &    &  45.296 \\
\enddata

\tablenotetext{a}{Blue luminosities in units of L$_\odot$}
\tablenotetext{b}{Observed H$_\beta$ luminosities in units of erg
s$^{-1}$.} \tablenotetext{c}{Extinction-corrected [O {\sc
iii}]$\lambda$5007 luminosities in units of erg s$^{-1}$.}
\tablenotetext{d}{Absorption-corrected 2-10 kev luminosities for
Compton-thin Seyfert 2 galaxies in units of erg s$^{-1}$.}
\tablenotetext{e}{Eddington luminosities in units of erg
s$^{-1}$.} \tablecomments{{\sc References}: a91=Acker et al. 1991;
a00=Awaki et al. 2000; b99=Bassani et al. 1999; c94=Cruz-Gonzalez
et al. 1994; c00=Collinge \& Brandt 2000; d85=Dahari 1985;
d88=Dahari \& De Robertis 1988; d92=de Grijp et al. 1992; h97=Ho
et al. 1997; l92=Lonsdale 1992; l01=Levenson et al. 2001;
lu01=Lumsden et al. 2001; m92=Moran et al. 1992; m94=Mulchaey et
al. 1994; m98=Maiolino et al. 1998; o94=Oliva et al. 1994;
p96=Polletta et al. 1996; p02=Almudena Prieto et al, 2002;
r99=Risaliti et al. 1999; r00=Risaliti et al. 2000; s89=Storchi
Bergmann \& Pastoriza 1989; s95=Storchi-Bergmann et al. 1995;
tar=tartarus.gsfc.nasa.gov; t91=Televich et al. 1991; t99=Tran et
al. 1999; v95=Veilleux et al. 1995; v97=Vaceli et al. 1997;
v02=Vanzi et al. 2002; w92=Whittle 1992; y96=Young et al. 1996.}
\end{deluxetable}

\clearpage
\begin{deluxetable}{lrrr}
  \tablewidth{0pt}
  \tablecaption{Statistical results for Seyfert 2 galaxies}
  \tablehead{
  \colhead{parameters} &  \colhead{Sy2s with PBL} &
  \colhead{Sy2s without PBL} & \colhead{possibility$^{a}$} }
\startdata
log$_{10}$N$_{\rm H}$ (cm$^{-2}$) & 23.633 $\pm$ 0.201 & 23.562 $\pm$ 0.350 & 79\%$^{c}$\\
log$_{10}$L$_{\rm B}$ (L$_\odot$)& 10.460 $\pm$ 0.068 & 10.340 $\pm$ 0.087 & 69\%$^{b}$ \\
log$_{10}$L$_{\rm FIR}$(L$_\odot$)& 10.506 $\pm$ 0.106 & 10.301$\pm$ 0.106 & 38\%$^{b}$\\
log$_{10}$(L$_{\rm FIR}$/L$_{\rm B}$) & -0.275 $\pm$ 0.069 & -0.298 $\pm$ 0.078 & 21\%$^{b}$\\
f$_{60}$/f$_{25}$ & 2.447 $\pm$ 0.201 & 4.974 $\pm$ 0.375 & 0.002\%$^{b}$\\
log$_{10}$L$_{25 \mu m}$ (L$_\odot$)& 10.565 $\pm$ 0.102 & 10.021 $\pm$ 0.111 & 0.005\%$^{b}$ \\
log$_{10}$L$_{\rm 1.49GHz}$\rm (erg s$^{-1}$ Hz$^{-1}$) & 29.803 $\pm$ 0.136 & 29.109 $\pm$ 0.135 & 0.02\%$^{b}$\\
log$_{10}$L$_{\rm [O III]}$\rm (erg s$^{-1}$) & 42.147 $\pm$ 0.123 & 41.426 $\pm$ 0.136 &  0.07\%$^{c}$ \\
log$_{10}$L$_{\rm 2-10kev}$\rm (erg s$^{-1}$)$^{d}$ & 42.996 $\pm$ 0.138 & 41.220 $\pm$ 0.320 & 0.001\%$^{c}$ \\
\enddata

\tablenotetext{a}{The possibility from the same parent
population.} \tablenotetext{b}{Two-sample Kolmogorov $-$ Smirnov
test.} \tablenotetext{c}{Survival data analysis program {\it
asurv} Rev 1.2, Gehan's Generalized Wilcoxon Test (Hypergeometric
variance).} \tablenotetext{d}{Just for Compton-thin Seyfert 2
galaxies.}
\end{deluxetable}

\clearpage
\begin{deluxetable}{ccccc}
  \tablewidth{0pt}
  \tablecaption{Common Sy2s observed in different samples}
  \tablehead{ \colhead{Samples $^{a}$} &
  \colhead{Y96 } &  \colhead{M00} &
  \colhead{T01 } & \colhead{L01 } }
\startdata
Y96 & 24(11)$^{b}$ & 2(1) & 9(9$^{c}$) & 3(3) \\
M00 &        & 31(11) & 13(5) & 4(1) \\
T01  &        &        & 49(22)& 16(8) \\
L01 &       &        &       & 22(8) \\
\enddata

\tablenotetext{a}{References for samples: Y96 = Young et al. 1996;
M00 = Moran et al., 2000; T01 = Tran 2001; L01 = Lumsden et al.
2001} \tablenotetext{b}{The numbers indicate the numbers of common
Seyfert 2 galaxies and the numbers in parentheses are numbers of
Sy2s with PBL} \tablenotetext{c}{Only 6 Sy2s with PBL have been
detected by Young et al. (1996).}
\end{deluxetable}
\end{document}